\documentclass[%
superscriptaddress,
reprint,
 amsmath,amssymb,
 aps,
prstab,
physics,
longbibliography,
]{revtex4-1}

\usepackage{graphicx}
\usepackage{dcolumn}
\usepackage{bm}
\usepackage{siunitx}
\usepackage{upgreek}

\usepackage[mathlines]{lineno}

\newcommand{\um}{\upmu \textrm{m}}

\begin{document}

\title{Acceleration of a Positron Bunch in a Hollow Channel Plasma}

\author{Spencer Gessner}
\email{sgess@slac.stanford.edu}
\affiliation{SLAC National Accelerator Laboratory, Menlo Park, California 94025, USA}
\author{Erik Adli} 
\affiliation{Department of Physics, University of Oslo, 0316 Oslo, Norway}
\author{James M. Allen}
\affiliation{SLAC National Accelerator Laboratory, Menlo Park, California 94025, USA}
\author{Weiming An}
\affiliation{Department of Astronomy, Beijing Normal University, Beijing 100875, China}
\author{Christine I. Clarke}
\affiliation{SLAC National Accelerator Laboratory, Menlo Park, California 94025, USA}
\author{Chris E. Clayton}
\affiliation{Department of Electrical Engineering, University of California Los Angeles, Los Angeles, California 90095, USA}
\author{Sebastien Corde}
\affiliation{LOA, ENSTA ParisTech, CNRS, Ecole Polytechnique, Universit\'e Paris-Saclay, 91762 Palaiseau, France}
\author{Antoine Doche}
\affiliation{LOA, ENSTA ParisTech, CNRS, Ecole Polytechnique, Universit\'e Paris-Saclay, 91762 Palaiseau, France}
\author{Joel Frederico}
\affiliation{SLAC National Accelerator Laboratory, Menlo Park, California 94025, USA}
\author{Selina Z. Green}
\affiliation{SLAC National Accelerator Laboratory, Menlo Park, California 94025, USA}
\author{Mark J. Hogan}
\affiliation{SLAC National Accelerator Laboratory, Menlo Park, California 94025, USA}
\author{Chan Joshi}
\affiliation{Department of Electrical Engineering, University of California Los Angeles, Los Angeles, California 90095, USA}
\author{Carl A. Lindstr\o m}
\affiliation{Department of Physics, University of Oslo, 0316 Oslo, Norway}
\author{Michael Litos}
\affiliation{University of Colorado Boulder, Boulder, CO 80309, USA}
\author{Kenneth A. Marsh}
\affiliation{Department of Electrical Engineering, University of California Los Angeles, Los Angeles, California 90095, USA}
\author{Warren B. Mori}
\affiliation{Department of Physics and Astronomy, University of California Los Angeles, Los Angeles, California 90095, USA}
\author{Brendan O'Shea}
\affiliation{SLAC National Accelerator Laboratory, Menlo Park, California 94025, USA}
\author{Navid Vafaei-Najafabadi}
\affiliation{Stonybrook University, Stony Brook, NY 11794, USA}
\author{Vitaly Yakimenko}
\affiliation{SLAC National Accelerator Laboratory, Menlo Park, California 94025, USA}

\date{\today}

\begin{abstract}
Plasmas are a compelling medium for particle acceleration owing to their natural ability to sustain large electric fields. Plasmas are also unique amongst accelerator technologies in that they respond differently to beams of opposite charge. The asymmetric response of a plasma to highly-relativistic electron and positron beams arises from the fact that plasmas are composed of light, mobile electrons and heavy, stationary ions. Hollow channel plasma acceleration is a technique for symmetrizing the response of the plasma, such that it works equally well for high-energy electron and positron beams. In the experiment described here, we demonstrate the generation of a positron beam-driven wake in an extended, annular plasma channel, and acceleration of a second trailing witness positron bunch by the wake. The leading bunch excites the plasma wakefield and loses energy to the plasma, while the witness bunch experiences an accelerating field and gains energy, thus providing a proof-of-concept for hollow channel acceleration of positron beams. At a bunch separation of 330 $\um$, the accelerating gradient is 70 MV/m, the transformer ratio is 0.55, and the energy transfer efficiency is 18\% for a drive-to-witness beam charge ratio of 5:1.
\end{abstract}

\maketitle

\section{\label{sec:intro}Introduction}

Plasma wakefield acceleration (PWFA) is a novel technology which can be used to accelerate particle beams with large gradients and high efficiency~\cite{Litos2014,Lindstrom2021a}. The most ambitious application of plasma accelerator technology is the generation of ultra-high energy, low-emittance beams for a plasma-based linear collider (PLC). There are several challenges on the path to the PLC. Chief among them is the acceleration of positron beams in plasma~\cite{Cao2023}. For linear collider designs, such as the ILC~\cite{ILC2022}, CLIC~\cite{CLIC2022}, and C$^3$~\cite{Vernieri2023}, which use radio-frequency (RF) waves to accelerate particles, electron and positron beams are accomodated by adjusting the phase of the RF wave. By contrast, the plasma accelerator reacts differently to beams of opposite charge. The plasma is composed of light, mobile electrons, and heavy, sluggish ions. When operating in the blowout regime~\cite{Lu2006}, an electron beam propagating into a plasma expels plasma electrons, creating a plasma bubble with strong ion focusing, which is advantageous for transporting and accelerating the electron beam, while maintaining the beam's emittance. On the other hand, a positron beam attracts plasma electrons toward the beam axis, which creates a complicated wakefield structure~\cite{Hogan2003,Blue2003,Muggli2008,Corde2015}. Recently, we have demonstrated that nonlinear wakes driven by positron beams in a uniform plasma can be used to accelerate trailing witness positron bunches with modest gradients but with high energy-extraction efficiency from the wake~\cite{Doche2017}. However, due to the complex focusing field structure of these wakes, the positron beam emittance, which is a measure of beam quality, is not preserved using this technique. In the moderately non-linear regime, it is possible to preserve the positron beam emittance in a uniform plasma, but this technique does not scale well to collider-quality emittances~\cite{Hue2021,Cao2023}. We are therefore motivated to pursue other approaches which in principle preserve the positron beam emittance. The hollow channel plasma wakefield accelerator is an appealing concept for accelerating positron beams in plasma because it can avoid the complicated wake structure that arises from non-linear plasma wakefields in uniform plasmas.

\begin{figure*}[]
\centering
\includegraphics[width=0.9\textwidth]{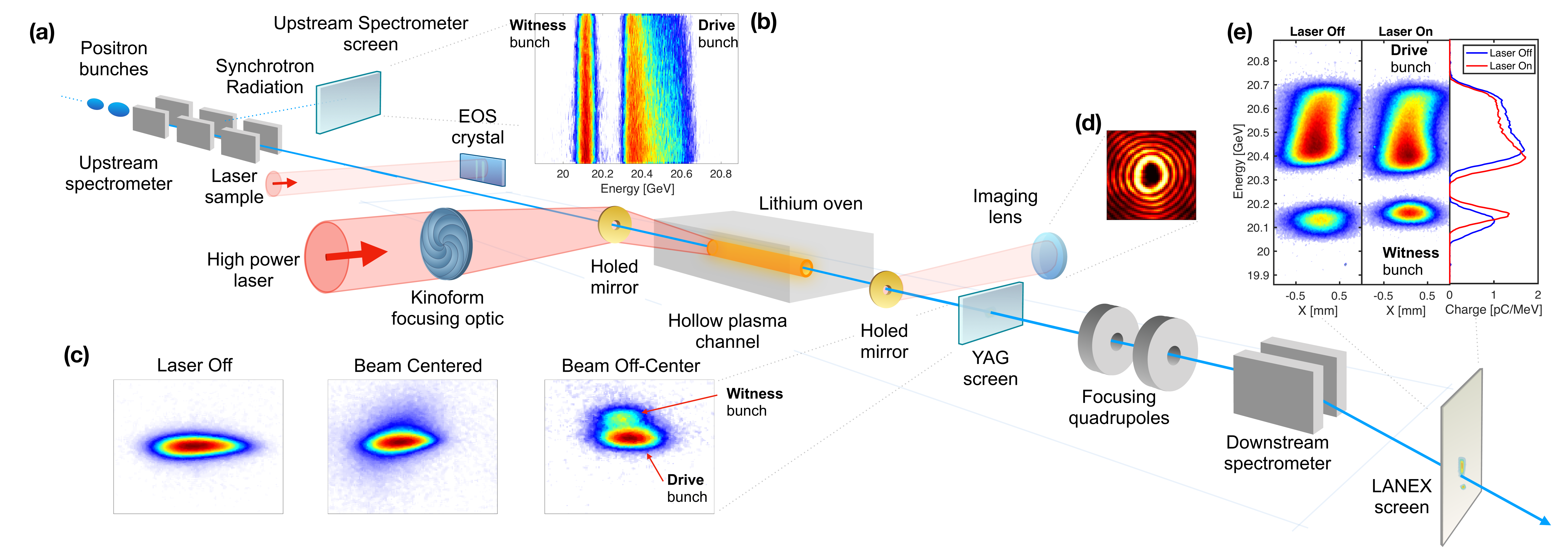}

\caption{(a). Schematic of the hollow channel experiment carried out at FACET. All relevant diagnostics are labeled. Inset (b) The drive-witness energy spectrum measured upstream of the hollow channel plasma. Inset (c) Sample images on the YAG screen downstream of the hollow channel plasma showing the transverse beam profile for different experimental conditions. Inset (d) Sample image of the ionizing high-order Bessel pulse used to ionize the plasma channel. Note that the image is deformed by the downstream optics and is primarily used for shot-to-shot jitter measurements. Inset (e) Comparison of beam energy spectra with and without the plasma channel present.} 
\label{fig:schem}
\end{figure*}

The hollow channel plasma wakefield accelerator was originally conceived as a technique for guiding high-intensity laser pulses for laser wakefield acceleration~\cite{Chiou1995}, and later recognized to have advantageous properties for beam-driven wakefield acceleration~\cite{Chiou1998}. In this scenario, an electron or positron beam propagates through a hollow tube of plasma and drives a high-amplitude electromagnetic field in its wake. The longitudinal field inside the channel is radially uniform, so that particles with different initial offsets with respect to the channel axis all gain energy at the same rate. The absence of plasma in the channel implies that there are no transverse forces from on-axis plasma electrons. Finally, for drive and witness beams propagating on-axis through the channel, there are no transverse forces and the beam emittance $\varepsilon$ is preserved through the acceleration process. However, for long acceleration lengths ($L \gg \beta_{x,y}$) an external focusing magnetic field may be necessary to guide both the drive and witness beam. Also, small perturbations of either the drive or witness beam centroids from the channel axis lead to the appearance of transverse wakefields that can seed and amplify the beam breakup instability~\cite{Schroeder1999} which has recently been observed experimentally~\cite{Lindstrm2018}. Additionally, asymmetries in the shape of the hollow plasma channel may lead to fields which deviate from the ideal scenario.

In previous work~\cite{Gessner2016}, we described the excitation of a hollow channel wakefield by a single positron bunch. Here we describe the subsequent acceleration of a witness positron bunch in the hollow channel wakefield. The hollow channel plasma is generated by ionizing a lithium vapor~\cite{Muggli1999} with a ring-shaped, high-intensity laser pulse using a phase plate that transforms an initial Gaussian beam into a high-order Bessel beam~\cite{Kimura2011}. The approximately annular plasma channel is produced by multi-photon ionization of lithium. We propagate a positron drive beam through the channel, creating a longitudinal wakefield that is used to accelerate a witness positron bunch. By scanning the separation of the drive and witness beams, we map the longitudinal shape of the wakefield and determine the accelerating phase~\cite{Schrder2020}. From the energy gained and lost by the witness and drive bunches, respectively, we deduce the net energy transfer efficiency and the transformer ratio at the phase corresponding to the maximum accelerating gradient.

\section{Experimental Overview}

The experiment was carried out at the Facility for Advanced aCcelerator Experimental Tests (FACET) at the SLAC National Accelerator Laboratory~\cite{Hogan2010}. SLAC is the only laboratory with the infrastructure required for providing high-energy, high-intensity positron beams for PWFA experiments. A low-emittance positron beam is extracted from the damping ring (see Appendix~\ref{meth:posGen}) and accelerated in the linac to 20 GeV energy. Along the linac, the particle beam is sent through a series of bunch compression chicanes which reduce the longitudinal size of the bunch $\sigma_z$ from several millimeters to a few hundred microns. 
In the final chicane, the beam is sent through a notch-collimation system~\cite{Litos2014} which exploits the time-energy correlation of the beam to convert the beam from a single-bunch structure to a two-bunch structure with variable separation. The collimation system is also used to adjust the total charge of the beam delivered to the experiment. At the end of the final chicane, the beam is focused by a series of quadrupole magnets before entering the lithium oven.

The schematic of the experiement is depicted in Figure~\ref{fig:schem}. The longitudinal shape of the two-bunch beam structure is characterized by an electro-optical sampling (EOS) crystal~\cite{Berden2007} with a resolution of 10 microns (30 femtoseconds). The plasma source is a lithium heat-pipe oven~\cite{Muggli1999}. The plasma source has several heating coils which can be turned on and off to adjust the length of the vapor region. The lithium is ionized by a Terawatt-class, Ti:Sapphire laser with 800 nm central wavelength. For this experiment, the laser is operated at low intensity with a maximum pulse energy of 40 mJ in a 75 fs-long pulse. In order to ionize the 25 cm-long annular plasma, the laser is shaped by a diffractive kinoform optic~\cite{Andreev1996,Fan2000,Kimura2011,Gessner2016}. The optic forms the laser into a high-order Bessel intensity profile with the first maximum occurring at a radius of $250~\um$ (see Appendix~\ref{meth:kino}). The Bessel intensity profile has the attractive feature that the transverse shape of the profile does not depend on the $z$-coordinate; the laser ionizes a channel with a fixed radius along the lithium oven. The optic is mounted on a 2D stage which allows for control of the transverse position of the laser focus.

The laser is directed onto the particle beam axis by a gold mirror with a hole for the positron beam to pass through. The laser pulse and positron beam co-propagate into the lithium vapor source. A delay stage is used to set the relative laser-positron timing and adjust the longitudinal position of the laser focus. The delay was set such that the laser arrives at most 5 picoseconds ahead of the positron beam, and with the laser focus set to the downstream end of the lithium vapor source, creating a 25$\pm$1 cm-long plasma channel. The energy of the ionized plasma electrons is less than 1 eV and have a velocity of $0.4~\um$/s. The motion of the plasma electrons has negligible effect on the shape of the channel prior to the arrival of the positron beam.

Downstream of the lithium oven, a second gold mirror with a hole deflects the laser pulse off of the beam axis. The spent laser pulse is re-imaged onto two cameras which we use to track the position of the laser focus during the experiment. The positron beam passes through the holed-mirror and through a yttrium aluminum garnet (YAG) crystal which is used to track the transverse position and size of the positron beam. Finally, the beam passes through a spectrometer dipole and the beam energy is measured on a scintillating LANEX screen (see Appendix~\ref{meth:spectrometer}).

\section{Results and Discussion}

\begin{figure}[tbh]
\centering
\includegraphics[width=\linewidth]{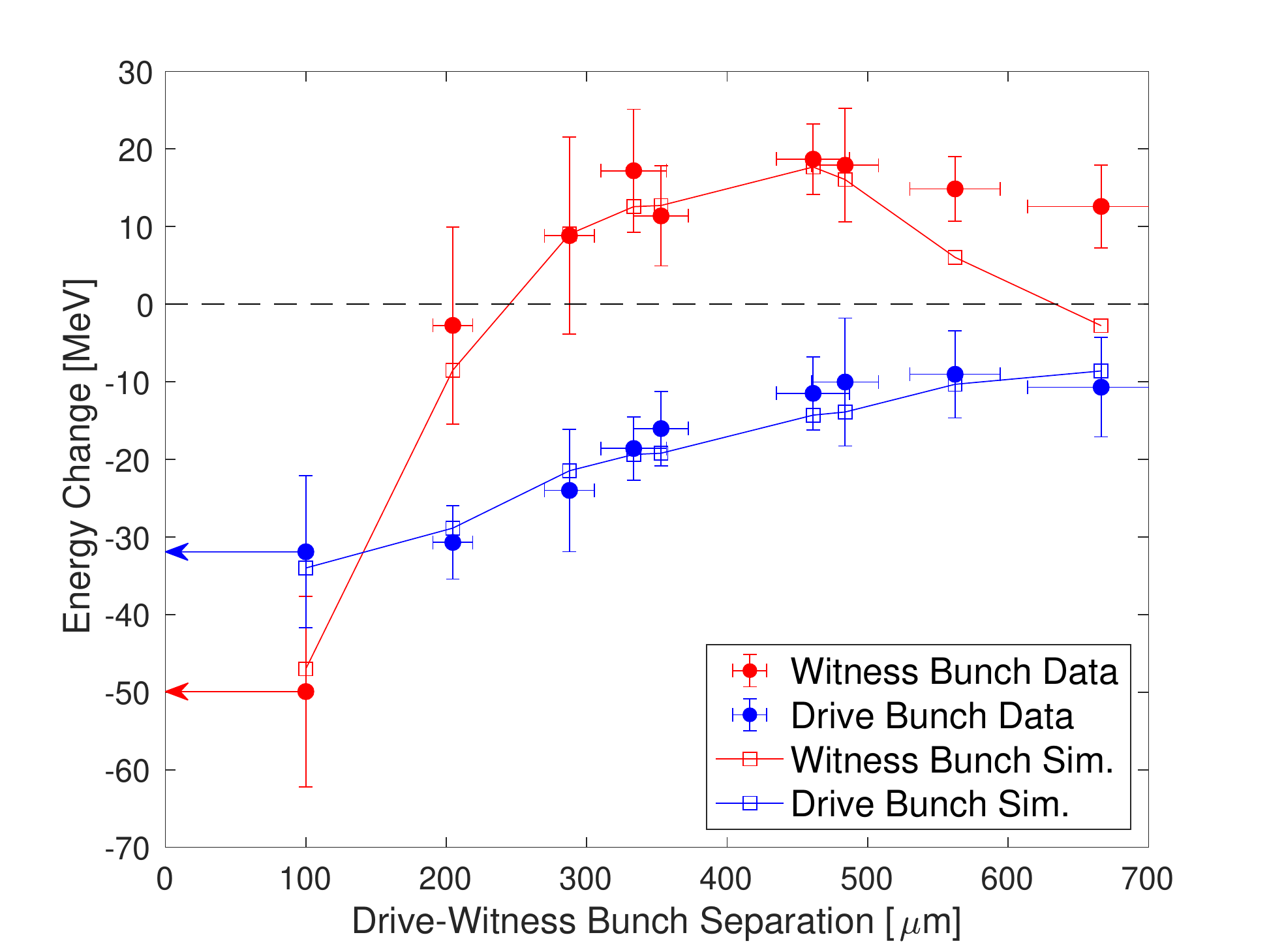}

\caption{Results of the longitudinal bunch separation scan and supporting QuickPIC simulation scan. Witness bunch data is denoted by red circles, and witness bunch simulation results shown with open red squares. Drive bunch data is denoted by blue circles, and drive bunch simulation results shown with open blue squares. For the first point in the scan, the drive and witness beams are not clearly distinguished and the data point is placed at an upper limit of 95 $\um$ bunch separation. Error bars represent the $1\sigma$ standard deviation of the measurement. The QuickPIC simulation use the parameters measured for each scan point described in Appendix~\ref{meth:sims}.  }
\label{fig:scan}
\end{figure}

In the first phase of the experiment, the transverse and longitudinal positron beam profiles and energy spectrum are measured for the beam propagating in vacuum. The lithium plasma source is heated to produce lithium vapor with a density of $3\times 10^{16}$ cm$^{-3}$. At high charge, the positron beam fields are large enough to self-ionize the neutral lithium atoms within the laser-produced hollow channel. The incoming positron charge was therefore reduced until there was no evidence of plasma interaction. The procedure was performed at peak bunch compression (maximum positron beam field strength). Table~\ref{tab:params} shows the positron beam parameters at the end of the charge reduction procedure. All parameters are measured at the interaction point, except for the beam emittance which was measured upstream of the final bunch compression chicane.

\begin{table}[]
    \centering
    \begin{tabular}{| c | c | c | c | c | c | c |}
    \hline
       $N_{tot}$ & $\sigma_x(\um$)  & $\sigma_y(\um$) & $\beta_x$(m) & $\beta_y$(m) & $\varepsilon_x(\um$) & $\varepsilon_y(\um$) \\
    \hline
       $3.5\times 10^9$ & 35 & 25 & 0.5 & 5.0 & 97 & 5 \\
    \hline
    \end{tabular}
    \caption{Positron beam parameters of the combined drive and witness bunches after the charge reduction procedure. All parameters are measured at the interaction point, except for the beam emittance which was measured upstream of the final bunch compression chicane.}
    \label{tab:params}
\end{table}

At peak compression, the EOS diagnostic does not clearly distinguish between the drive and witness bunches, but can be used to set an upper limit on the longitudinal separation between the two bunches $\Delta_{dw} < 95~\um$ . The drive and witness bunches are well separated on the spectrometer screen, as seen in Figure~\ref{fig:schem}b), with drive bunch energy $E_d = 20.487$ GeV and witness bunch energy $E_w = 20.095$ GeV. The energy resolution of the spectrometer is 3 MeV.

After confirming non-interaction between the beam and lithium vapor, the laser was fired to generate the hollow channel plasma. The laser profile was previously optimized according to the procedure described in Appendix~\ref{meth:chanOpt}. Next, we performed a longitudinal drive-witness separation scan to map out the hollow channel plasma wakefield, according to the procedure described in Appendix~\ref{meth:scans}. The drive-witness separation is measured by the EOS system on every shot and the spectrometer measures the change in energy of the drive and witness beams. 

Figure~\ref{fig:scan} shows the results of the scan. In the ideal bunch separation scan, all beam parameters would be kept constant as the bunch separation is varied. In practice, the bunch charge, length, and separation are coupled together and depend on the longitudinal phase space. We actively corrected the bunch charge and beam trajectory throughout the scan, but were not able to control the length of the drive bunch independently of the bunch separation. The length of the drive bunch increases by a factor of five over the course of the scan while the drive charge varies by $\pm 18\%$. The witness bunch charge and length are stable at the level of $\pm 20\%$. We observed that the magnitude of the energy loss experienced by the drive bunch decreased over the course of the scan, which is consistent with increasing drive bunch length. From the maximum energy gain of 17.5 MeV at a bunch separation of 330 $\um$, we infer an accelerating gradient of $70\pm30$ MV/m, and the uncertainty is the statistical $1\sigma$ standard deviatio of the measurement.

\begin{figure}[tbh]
\centering
\includegraphics[width=\linewidth]{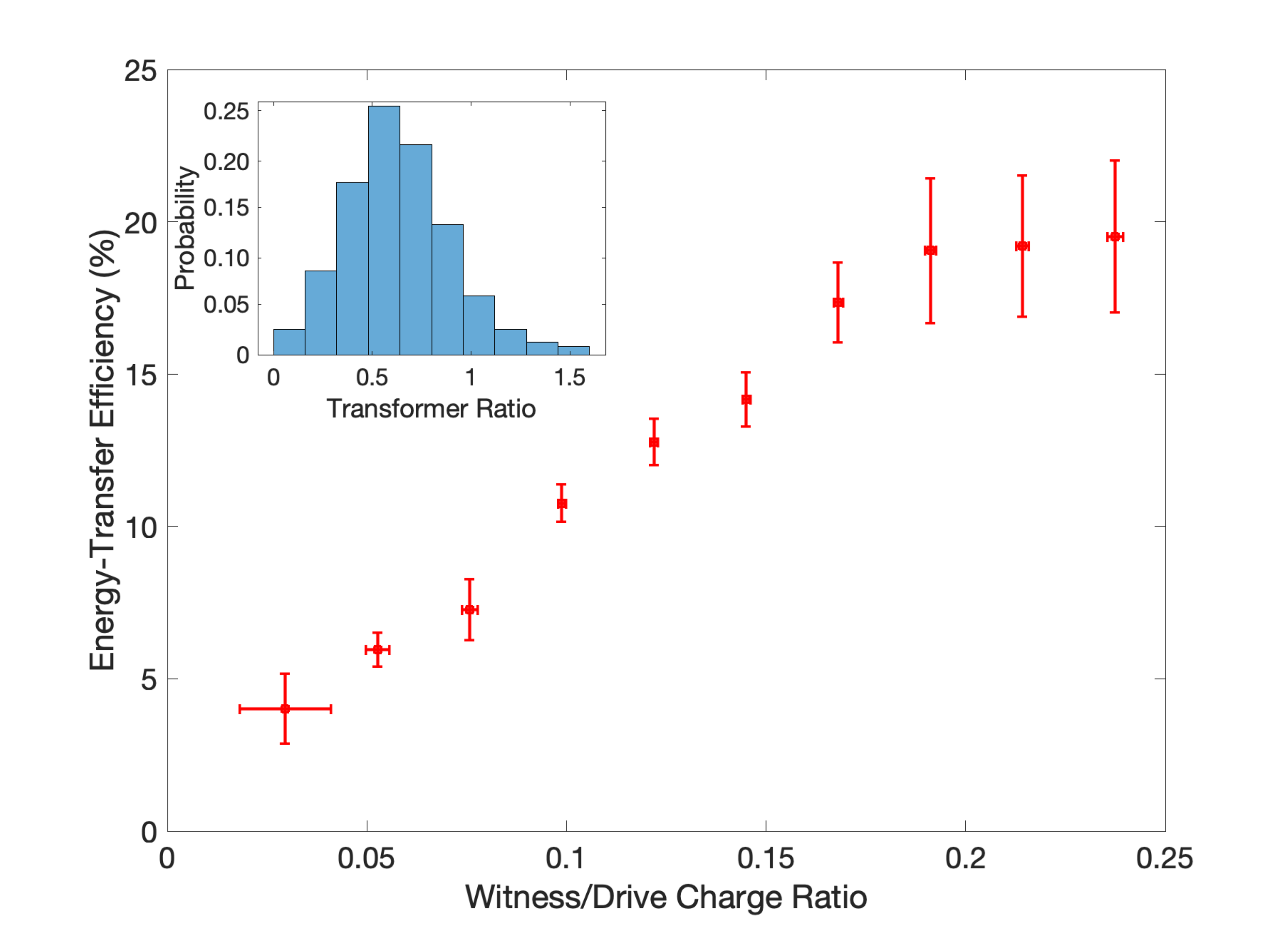}
\caption{Efficiency of energy transfer from drive bunch to witness bunch for the $\Delta_{dw} = 330~\um$ scan point as a function of witness-to-drive charge ratio. The error bars denote the error on the mean of the values in a given witness-to-drive charge ratio bin due to jitter in the measurement. Inset: Normalized histogram of transformer ratios for the same set of shots.}
\label{fig:trans}
\end{figure}

We analyzed the efficiency and transformer ratio of the wakefield at a bunch separation of 330 $\um$. The transformer ratio is defined as 
\begin{equation}
    T = \left|\frac{\max(E_z)}{\min(E_z)}\right|.
\end{equation}
While we do not measure the accelerating field directly, the centroid energy gain of the witness bunch is a good proxy for the peak field because the witness bunch is much shorter than the wavelength of the wakefield. Measuring the peak decelerating field is complicated by the fact that the drive bunch experiences a changing field over its length. We can extract the peak decelerating field from the drive bunch spectrum if the field is single-valued (monotonically decreasing) over the length of the bunch~\cite{Clayton2016}. This condition is satisfied for the scan step at 330 $\um$ separation because the $\sigma_{zd} = 122~\um$ and the end of the drive bunch sits in the peak decelerating phase of the wake. We exploit the linear correlation between position and energy in the drive beam phase space and measure the change in energy of the lower end of the spectrum to determine the peak decelerating field. By comparison, efficiency is a simpler quantity to measure
\begin{equation}
    \eta = \frac{\Delta E_w Q_w}{\Delta E_d Q_d},
\end{equation}
where $\Delta E_w$ and $\Delta E_d$ are the centroid change in energy of the witness and drive bunches, respectively. Figure~\ref{fig:trans} shows the efficiency of energy transfer from the drive beam to the witness beam as a function of the ratio of drive charge to witness charge. There is large shot-to-shot variation of the charge in the drive and witness bunches and the data is binned by the charge ratio. A histogram of measured transformer ratio values is shown as an inset, with ratios in excess of one observed on some shots. The median observed transformer ratio was 0.55. The efficiencies measured in this experiment are roughly a factor of 2 below the best efficiencies observed for plasma acceleration in a uniform plasma~\cite{Lindstrom2021a}.

\subsection{Comparison with Theory and Simulation}

\begin{figure*}[tbh]
\centering
\includegraphics[width=\linewidth]{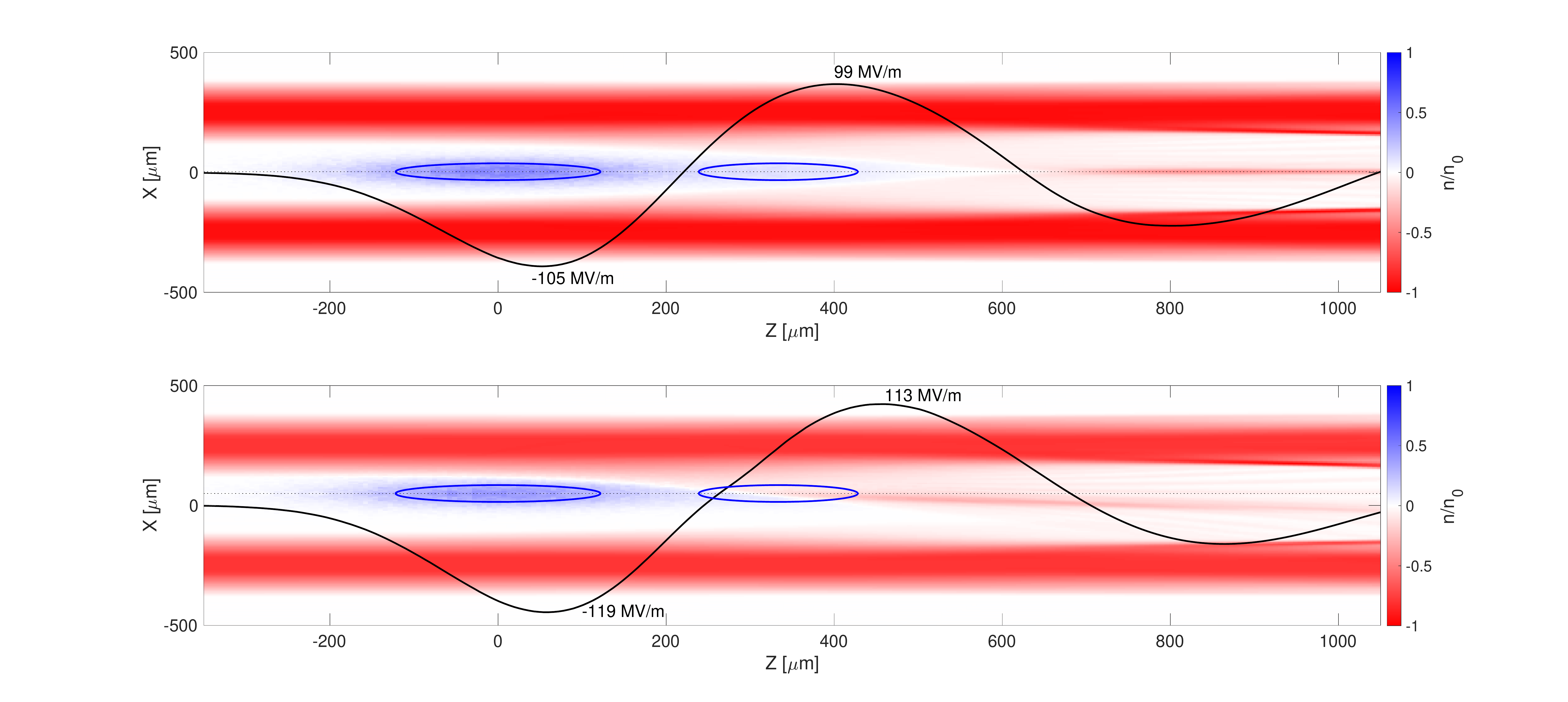}

\caption{QuickPIC simulations of positron drive and witness beams propagating in a hollow channel plasma channel. The beams propagate to the left. The blue circles denote the one-sigma beam contours. The dotted black line denotes the transverse beam offset. The solid black line is the accelerating field sampled at the transverse position of the beam. The maximum and minimum longitudinal field values are denoted on the field curves. Top: Both beams centered on axis. Bottom:  The drive and witness beams are offset from the axis by $65~\um$. The accelerating field is  enhanced and the wavelength is elongated for the case of the offset beam. Shot-to-shot jitter of the channel axis with respect to the beam axis is approximately $25~\um$ at the downstream end of the channel.  }
\label{fig:sims}
\end{figure*}

The wavelength of a hollow channel plasma wakefield is given by $\lambda = 2\pi/\chi_{||}k_p$ with $k_p$ the plasma wavenumber. The geometric factor $\chi_{||}$ is derived from the boundary conditions and given by~\cite{Gessner2016Thesis}
\begin{equation}\label{eq:chi}
    \chi_{||} = \sqrt{\frac{2 B_{10}(a,b)}{2B_{10}(a,b)-k_paB_{00}(a,b)}},
\end{equation}
with $B_{ij}$ the Bessel boundary function given by
\begin{equation}
    B_{i,j}(a,b) = I_i(k_p a) K_j (k_p b) + (-1)^{i-j+1}I_j(k_p b)K_i(k_p a),
\end{equation}
for inner plasma radius $a$ and outer plasma radius $b$. These equations are derived under the assumption that there is a sharp transition between the ionized and un-ionized regions. 

For fully-ionized vapor at $n_0=3\times10^{16}$ cm$^{-3}$, we expect to observe the accelerating phase of the wake for a bunch separation $\Delta_{dw} = 165~\um$, but in the scan, the peak accelerating phase is observed at $\Delta_{dw} = 460~\um$.  The bulk of the discrepancy is the result of reducing the laser intensity used to ionize the hollow channel plasma according to the channel symmetrization procedure described in Appendix~\ref{meth:chanOpt}. We use a multi-photon ionization model~\cite{PPT1966} to estimate that the ionized fraction of the vapor is roughly 10\%, which is consistent with the measurement and a rough scaling of the hollow channel plasma wavelength $\lambda \propto n^{-1/2}$, modulo the factor $\chi_{||}$ in Equation~\ref{eq:chi}.

The sharp-boundary model estimates that the peak accelerating phase should occur at a bunch separation of $\Delta_{dw} = 386~\um$, which is consistent with our observations. The model also predicts a return to the decelerating phase at large bunch separations, but the data show a long tapering-off of the energy gain at large bunch separation. We investigated a variety of possibilities, including the presence of a small amount of plasma inside the plasma channel~\cite{Schroeder2013}, but the best fit to the data is achieved by modeling the channel wall with a gradual density transition from the central unionized region to the fully-ionized region. We refer to this configuration as the ``soft boundary" model. We chose a linear ramp function for our model (see Appendix~\ref{meth:sims}) with a width of one skin depth. The exact shape of the transition from unionized inner region to fully-ionized outer region is not important. Rather, it is the length scale over which the transition occurs that is significant~\cite{Shvets1996}. In the sharp boundary model, the walls of the plasma channel contain the electromagnetic wake within the channel. In the soft boundary case, the electromagnetic wake interacts with plasma at a range of densities and frequencies over the transition region. This causes phase mixing and electrostatic separation of plasma electrons and ions. These effects both damp and lengthen the electromagnetic wake. Finally, the inclusion of transverse offsets of the drive and witness beam in the channel contributes to an even faster phase-mixing and wake-damping process because the outer edges of the beam propagate in the transition region. 

Figure~\ref{fig:sims} shows positron drive and witness beams propagating through a hollow channel plasma with a linear transition between the unionized and ionized regions. In the bottom simulation, the beams are offset from the axis by $65~\um$ and which contributes to damping and lengthening of the wake as compared to the on-axis case shown above. By including the soft boundary model and transverse offsets, we find good agreement between data and simulation for bunch separations up to $500~\um$, as shown in Figure~\ref{fig:scan}.

\section{Conclusion}

We have demonstrated the acceleration of a trailing positron bunch by a positron beam-driven wake in the hollow channel plasma. In this experiment, at a bunch separation of 330 $\um$, and a drive-to-witness charge ratio of 5:1, we observe an accelerating gradient of $70\pm30$ MV/m, a median transformer ratio of 0.55, and an energy transfer eﬃciency of 18\%. This important proof-of-concept shows the promise of hollow channel acceleration for positron beams, while also highlighting the engineering and physics challenges that must be overcome to further advance this technology. The most critical challenge is mitigation of the beam breakup instability~\cite{Lindstrm2018} that will dilute the emittance of the accelerating beam. The wake damping features of the soft boundary model affect both the $m=0$ longitudinal mode and $m=1$ dipole mode. Further studies are warranted to see if the $m=1$ mode can be preferentially damped over the $m=0$ mode through a judicious choice of transverse ramp parameters. 

The acceleration gradient in this experiment was limited by the requirement that the positron beam not self-ionize the residual neutral Li vapor inside the hollow plasma channel. There are several approaches that would allow experiments to reach GeV/m gradients in the hollow channel plasma. First, by switching to a different gas species with a higher ionization threshold, such as He, we will be able to increase the charge of the driving bunch without ionizing the residual vapor in the channel. Without changing any other aspects of the experiment, the accelerating gradient would increase to 300 MeV/m by increasing the drive charge to 2 nC. Next, the accelerating gradient will be increased by reducing the inner radius of the channel, which can be achieved by selecting a different kinoform optic. A narrower channel will also require greater beam stability from the accelerator in order to avoid excitation on transverse modes in the channel. An alternative method for creating the plasma channel may envoke self-guiding laser techniques~\cite{Feder2020}, but this would result in a not-quite hollow channel plasma similar to the scenario described by Schroeder~\cite{Schroeder2013}. 

Future plasma-based positron acceleration experiments will use electron beam drivers. The possibility of using electron beams to drive a wake with a  positron witness beam is a planned capability of FACET-II~\cite{Yakimenko2019}. Recent work on hollow channel plasma wakefield acceleration suggests that efficient beam loading can be achieved using an asymmetric drive electron beam that mitigates the beam break up instability of the positron bunch~\cite{Zhou2021}. Positron plasma acceleration remains a critical challenge on the path to a Plasma Linear Collider, and hollow channel plasmas have promising capabilities that require further development to be fully exploited.

\section{Acknowledgements}
This work is supported by the U.S. Department of Energy under Contract DE-
AC02-76SF0051 and National Natural Science Foundation of China (NSFC) Grant Nos. 12075030.

\appendix

\begin{table*}[]
    \centering
    \begin{tabular}{| c| c | c | c | c | c | c |}
    \hline
       Step & ~~$Q_d~(1\times 10^9)$~~ & ~~$\sigma_d~(\um$)~~ & ~~$Q_w~(1\times 10^9)$~~ & ~~$\sigma_w~(\um$)~~ & ~~$\Delta_{dw}~(\um$)~~ & ~~$\Delta_r~(\um$)~~ \\
       \hline
       1    & 2.72                 & 42                & 0.95                 & 42                & 98                   & 0                 \\
       2    & 2.92                 & 67                & 0.55                 & 65                & 205                  & 45                \\
       3    & 3.02                 & 107               & 0.59                 & 67                & 288                  & 66                \\
       4    & 2.91                 & 122               & 0.52                 & 95                & 333                  & 65                \\
       5    & 2.48                 & 119               & 0.85                 & 62                & 353                  & 78                \\
       6    & 2.57                 & 153               & 0.69                 & 71                & 461                  & 80                \\
       7    & 2.26                 & 153               & 0.74                 & 64                & 483                  & 90                \\
       8    & 2.20                 & 185               & 0.83                 & 58                & 562                  & 87                \\
       9    & 2.12                 & 208               & 0.90                 & 84                & 666                  & 77                \\
       \hline
    \end{tabular}
    \caption{Inputs to the QuickPIC simulations of the longitudinal beam separation scan. $\Delta_r$ is the offset between the beam and channel access.}
    \label{tab:quickpic}
\end{table*}

\section{\label{meth:posGen}Positron Beam Generation}
A 20 GeV electron beam is accelerated along a 2 kilometer-long S-band linac. The positron beam is generated by colliding the electron beam with a tungsten target. This produces a shower of high-emittance positrons that are captured in a solenoidal horn and transported back to the start of the linac where they are fed into a damping ring. The damping ring reduces the emittance of the positron beam to roughly 50 mm mrad. After damping, it is injected into the linac and accelerated to 20 GeV~\cite{Clendenin:1988np}.

\section{\label{meth:kino}Hollow Laser Optics}
The kinoform is a 1 mm thick piece of fused silica with an etched pattern that approximates the spiral phase $\Phi = k_\perp r + m\phi$ that imprints a high-order Bessel profile onto the laser pulse. Here, $r$ and $\phi$ are the radial and azimuthal coordinates, $m$ is the Bessel order, and $k_\perp = \gamma k$ with $\gamma$ the angle of focused rays with respect to the axis and $k$ is the wavenumber for 800 nm light. We chose $m=7$ and $\gamma = 4.4$ mrad. This produces a laser intensity profile of the form $I(r,z) = I_0 2\pi k z \gamma^2 J_7^2(k_\perp r)$ with the first maximum occurring at a radius of 250 $\um$~\cite{Gessner2016}. Here, $I_0$ is the incident laser intensity on the kinoform, $z$ is the longitudinal coordinate of laser propagation and $J_7$ is the seventh-order Bessel function of the first kind.

\section{\label{meth:chanOpt}Channel Optimization}
At the start of the experiment, the laser pulse used to create the plasma channel is  optimized for symmetry. Asymmetries may arise from phase errors that accumulate during laser transport or nonuniform illumination of the kinoform optic. Astigmatism is the most prominent asymmetry. For diagnostic purposes, we reduce the intensity of the laser in order to view the focal spot of the laser on a titanium foil just upstream of the plasma source. We adjust a lens in the transport system to remove astigmatism and optimize for uniformity at the focal point. This procedure is performed at low laser intensity to avoid burning the titanium foil. We then remove the titanium foils and increase the laser intensity to ionize lithium vapor.

We use a single positron beam to study the transverse shape of the hollow channel plasma. Since we cannot view the laser focus directly when operating with intensities large enough to ionize the plasma, we use beam-based measurements to characterize the shape of the Bessel focus. We use the 2D kinoform stage to scan the transverse position of the laser relative to the positron beam axis, while monitoring the position of the laser focus using the spent laser profiles. When the Bessel focus is set such that the beam propagates inside the channel, but slightly off-axis, the beam feels a transverse kick towards the wall of the plasma channel~\cite{Lindstrm2018}. We measure the deflection of the beam on the downstream YAG crystal. If the focus is set such that the positron beam is propagating in the ionized region, the beam experiences strong transverse focusing forces, but no deflection. In this case, we measure an increase in the beam size on the YAG screen due to the larger beam divergence. The scan reveals the center of hollow channel, as well as any asymmetries that may be present. If the scan reveals large asymmetries, we readjust the lens in the transport system and repeat the scan.

\section{\label{meth:scans}Longitudinal Phase Scans}
In the longitudinal phase scan, the separation between the drive and witness bunches is changed by adjusting the RF phase of the linac and by setting the position of the notch-collimation system. The scan requires multiple parameters to be tuned simultaneously, which affects the incoming energy distribution of the drive and witness beams. Changes to the beam energy affect the steering through the final bunch compression chicane, and that can lead to mis-steering or mis-focusing of the beams if the accelerator lattice is not compensated. During the longitudinal phase scan, the beam trajectory through the IP area is monitored and adjusted as needed. Despite best efforts, there is a net drift of the combined drive-witness transverse centroid of up to $90~\um$ over the course of the scan.

\section{\label{meth:spectrometer}Measuring Changes in the Beam Energy}
The dumpline energy spectrometer is composed of a quadrupole doublet and a dipole that bends the beam in the vertical plane. The doublet provides point-to-point imaging of the end of the plasma to the LANEX screen where it is viewed by a scientific-CMOS PCO Edge camera. The energy-per-pixel resolution of the camera is 2.9 MeV for the lower-energy witness bunch and 3.0 MeV at the higher-energy drive bunch. The calibration does not vary significantly over the individual bunches.

With the quadrupole doublet imaging the exit plane of the plasma, there are two first-order contributions to the position measured at the spectrometer screen: the change in beam energy $\Delta E$ and the beam offset from nominal trajectory at the end of the plasma $\Delta y_{plas}$. The change in the position of the beam on the screen due to the plasma interaction is given by
\begin{equation}
    \Delta y_{screen} = \eta(\Delta E/E_0) + R_{33}\Delta y_{plas},
\end{equation}
where $\eta$ is the dispersion due to the spectrometer dipole and $R_{33}$ is the magnification of the position of the beam at the exit of the plasma channel. With the laser off (no plasma interaction), we measure $R_{33} = 2.35$. A transverse vertical offset of $6~\um$ in the image plane is on the same scale as a 3 MeV energy change, and we therefore need to account for beam orbit changes when measuring the energy.

When correcting for orbit changes, we are faced with the issue that the BPMs do not distinguish between the drive and witness bunches. The YAG profile monitor downstream of the plasma can in some instances distinguish the two bunches, but not for the 5:1 drive:witness charge ratio used in this dataset. We therefore assume that both the drive bunch and witness bunch have the same offset as measured on the YAG screen and subtract off the orbital component from the beam energy measurement.

\section{\label{meth:sims}Simulations}
Simulations are performed with QuickPIC, a quasi-static Particle-in-Cell code~\cite{An2013}. The simulations are performed on a 3D grid with $256\times256\times256$ cells and a box size of $15k_p^{-1}\approx1500~\um$ in each dimension. The plasma channel walls are modeled as a trapezoidal function of the radius given by
\begin{equation}
    n_p(r) = 
    \begin{cases} 
    0, & r < a \\ 
    n_{0}\frac{r-a}{b-a}, & a \leq r < b \\
    n_{0}, & b \leq r < c \\
    n_{0}\frac{d - r}{d-c}, & c \leq r < d \\
    0, & d \leq r
    \end{cases}
\end{equation}
with $n_0 = 3\times10^{15}$ cm$^{-3}$, $a = 115~\um$, $b = 215~\um$, $c = 280~\um$, and $d = 380~\um$.

We used the beam parameters measured upstream of the hollow channel plasma interaction as inputs to QuickPIC to simulate the longitudinal separation scan. The inputs are shown in Table~\ref{tab:quickpic}. For all points in the scan, both the drive and witness beams have transverse size $\sigma_x = 35~\um$ and $\sigma_y = 25~\um$. Each point in the scan is a single-step simulation using a non-evolving beam. The energy gain/loss of the drive/witness beams are evaluated at the longitudinal centroid of the beams. When the beams have a transverse offset from the propagation axis, the field is sampled at the transverse centroid of the beams.

\bibliographystyle{apsrev}
\bibliography{main}

\begin{thebibliography}{35}
\expandafter\ifx\csname natexlab\endcsname\relax\def\natexlab#1{#1}\fi
\expandafter\ifx\csname bibnamefont\endcsname\relax
  \def\bibnamefont#1{#1}\fi
\expandafter\ifx\csname bibfnamefont\endcsname\relax
  \def\bibfnamefont#1{#1}\fi
\expandafter\ifx\csname citenamefont\endcsname\relax
  \def\citenamefont#1{#1}\fi
\expandafter\ifx\csname url\endcsname\relax
  \def\url#1{\texttt{#1}}\fi
\expandafter\ifx\csname urlprefix\endcsname\relax\def\urlprefix{URL }\fi
\providecommand{\bibinfo}[2]{#2}
\providecommand{\eprint}[2][]{\url{#2}}

\bibitem[{\citenamefont{Litos et~al.}(2014)\citenamefont{Litos, Adli, An, Clarke, Clayton, Corde, Delahaye, England, Fisher, Frederico et~al.}}]{Litos2014}
\bibinfo{author}{\bibfnamefont{M.}~\bibnamefont{Litos}}, \bibinfo{author}{\bibfnamefont{E.}~\bibnamefont{Adli}}, \bibinfo{author}{\bibfnamefont{W.}~\bibnamefont{An}}, \bibinfo{author}{\bibfnamefont{C.~I.} \bibnamefont{Clarke}}, \bibinfo{author}{\bibfnamefont{C.~E.} \bibnamefont{Clayton}}, \bibinfo{author}{\bibfnamefont{S.}~\bibnamefont{Corde}}, \bibinfo{author}{\bibfnamefont{J.~P.} \bibnamefont{Delahaye}}, \bibinfo{author}{\bibfnamefont{R.~J.} \bibnamefont{England}}, \bibinfo{author}{\bibfnamefont{A.~S.} \bibnamefont{Fisher}}, \bibinfo{author}{\bibfnamefont{J.}~\bibnamefont{Frederico}}, \bibnamefont{et~al.}, \bibinfo{journal}{Nature} \textbf{\bibinfo{volume}{515}}, \bibinfo{pages}{92} (\bibinfo{year}{2014}), \urlprefix\url{https://doi.org/10.1038/nature13882}.

\bibitem[{\citenamefont{Lindstr\o{}m et~al.}(2021)}]{Lindstrom2021a}
\bibinfo{author}{\bibfnamefont{C.~A.} \bibnamefont{Lindstr\o{}m}} \bibnamefont{et~al.}, \bibinfo{journal}{Phys. Rev. Lett.} \textbf{\bibinfo{volume}{126}}, \bibinfo{pages}{014801} (\bibinfo{year}{2021}), \urlprefix\url{https://doi.org/10.1103/PhysRevLett.126.014801}.

\bibitem[{\citenamefont{Cao et~al.}(2023)\citenamefont{Cao, Lindstrøm, Adli, Corde, and Gessner}}]{Cao2023}
\bibinfo{author}{\bibfnamefont{G.~J.} \bibnamefont{Cao}}, \bibinfo{author}{\bibfnamefont{C.~A.} \bibnamefont{Lindstrøm}}, \bibinfo{author}{\bibfnamefont{E.}~\bibnamefont{Adli}}, \bibinfo{author}{\bibfnamefont{S.}~\bibnamefont{Corde}}, \bibnamefont{and} \bibinfo{author}{\bibfnamefont{S.}~\bibnamefont{Gessner}}, \emph{\bibinfo{title}{Positron acceleration in plasma wakefields}} (\bibinfo{year}{2023}), \urlprefix\url{https://arxiv.org/abs/2309.10495}.

\bibitem[{\citenamefont{Aryshev et~al.}(2022)}]{ILC2022}
\bibinfo{author}{\bibfnamefont{A.}~\bibnamefont{Aryshev}} \bibnamefont{et~al.}, \emph{\bibinfo{title}{The international linear collider: Report to snowmass 2021}} (\bibinfo{year}{2022}), \urlprefix\url{https://arxiv.org/abs/2203.07622}.

\bibitem[{\citenamefont{Brunner et~al.}(2022)\citenamefont{Brunner, Burrows, Calatroni, Lasheras, Corsini, D'Auria, Doebert, Faus-Golfe, Grudiev, Latina et~al.}}]{CLIC2022}
\bibinfo{author}{\bibfnamefont{O.}~\bibnamefont{Brunner}}, \bibinfo{author}{\bibfnamefont{P.~N.} \bibnamefont{Burrows}}, \bibinfo{author}{\bibfnamefont{S.}~\bibnamefont{Calatroni}}, \bibinfo{author}{\bibfnamefont{N.~C.} \bibnamefont{Lasheras}}, \bibinfo{author}{\bibfnamefont{R.}~\bibnamefont{Corsini}}, \bibinfo{author}{\bibfnamefont{G.}~\bibnamefont{D'Auria}}, \bibinfo{author}{\bibfnamefont{S.}~\bibnamefont{Doebert}}, \bibinfo{author}{\bibfnamefont{A.}~\bibnamefont{Faus-Golfe}}, \bibinfo{author}{\bibfnamefont{A.}~\bibnamefont{Grudiev}}, \bibinfo{author}{\bibfnamefont{A.}~\bibnamefont{Latina}}, \bibnamefont{et~al.}, \emph{\bibinfo{title}{The clic project}} (\bibinfo{year}{2022}), \urlprefix\url{https://arxiv.org/abs/2203.09186}.

\bibitem[{\citenamefont{Vernieri et~al.}(2023)\citenamefont{Vernieri, Nanni, Dasu, Peskin, Barklow, Bartoldus, Bhat, Black, Brau, Breidenbach et~al.}}]{Vernieri2023}
\bibinfo{author}{\bibfnamefont{C.}~\bibnamefont{Vernieri}}, \bibinfo{author}{\bibfnamefont{E.~A.} \bibnamefont{Nanni}}, \bibinfo{author}{\bibfnamefont{S.}~\bibnamefont{Dasu}}, \bibinfo{author}{\bibfnamefont{M.~E.} \bibnamefont{Peskin}}, \bibinfo{author}{\bibfnamefont{T.}~\bibnamefont{Barklow}}, \bibinfo{author}{\bibfnamefont{R.}~\bibnamefont{Bartoldus}}, \bibinfo{author}{\bibfnamefont{P.~C.} \bibnamefont{Bhat}}, \bibinfo{author}{\bibfnamefont{K.}~\bibnamefont{Black}}, \bibinfo{author}{\bibfnamefont{J.~E.} \bibnamefont{Brau}}, \bibinfo{author}{\bibfnamefont{M.}~\bibnamefont{Breidenbach}}, \bibnamefont{et~al.}, \bibinfo{journal}{Journal of Instrumentation} \textbf{\bibinfo{volume}{18}}, \bibinfo{pages}{P07053} (\bibinfo{year}{2023}), ISSN \bibinfo{issn}{1748-0221}, \urlprefix\url{http://dx.doi.org/10.1088/1748-0221/18/07/P07053}.

\bibitem[{\citenamefont{Lu et~al.}(2006)\citenamefont{Lu, Huang, Zhou, Mori, and Katsouleas}}]{Lu2006}
\bibinfo{author}{\bibfnamefont{W.}~\bibnamefont{Lu}}, \bibinfo{author}{\bibfnamefont{C.}~\bibnamefont{Huang}}, \bibinfo{author}{\bibfnamefont{M.}~\bibnamefont{Zhou}}, \bibinfo{author}{\bibfnamefont{W.~B.} \bibnamefont{Mori}}, \bibnamefont{and} \bibinfo{author}{\bibfnamefont{T.}~\bibnamefont{Katsouleas}}, \bibinfo{journal}{Physical Review Letters} \textbf{\bibinfo{volume}{96}} (\bibinfo{year}{2006}), \urlprefix\url{https://doi.org/10.1103/physrevlett.96.165002}.

\bibitem[{\citenamefont{Hogan et~al.}(2003)\citenamefont{Hogan, Clayton, Huang, Muggli, Wang, Blue, Walz, Marsh, O'Connell, Lee et~al.}}]{Hogan2003}
\bibinfo{author}{\bibfnamefont{M.~J.} \bibnamefont{Hogan}}, \bibinfo{author}{\bibfnamefont{C.~E.} \bibnamefont{Clayton}}, \bibinfo{author}{\bibfnamefont{C.}~\bibnamefont{Huang}}, \bibinfo{author}{\bibfnamefont{P.}~\bibnamefont{Muggli}}, \bibinfo{author}{\bibfnamefont{S.}~\bibnamefont{Wang}}, \bibinfo{author}{\bibfnamefont{B.~E.} \bibnamefont{Blue}}, \bibinfo{author}{\bibfnamefont{D.}~\bibnamefont{Walz}}, \bibinfo{author}{\bibfnamefont{K.~A.} \bibnamefont{Marsh}}, \bibinfo{author}{\bibfnamefont{C.~L.} \bibnamefont{O'Connell}}, \bibinfo{author}{\bibfnamefont{S.}~\bibnamefont{Lee}}, \bibnamefont{et~al.}, \bibinfo{journal}{Physical Review Letters} \textbf{\bibinfo{volume}{90}} (\bibinfo{year}{2003}), \urlprefix\url{https://doi.org/10.1103/physrevlett.90.205002}.

\bibitem[{\citenamefont{Blue et~al.}(2003)\citenamefont{Blue, Clayton, O'Connell, Decker, Hogan, Huang, Iverson, Joshi, Katsouleas, Lu et~al.}}]{Blue2003}
\bibinfo{author}{\bibfnamefont{B.~E.} \bibnamefont{Blue}}, \bibinfo{author}{\bibfnamefont{C.~E.} \bibnamefont{Clayton}}, \bibinfo{author}{\bibfnamefont{C.~L.} \bibnamefont{O'Connell}}, \bibinfo{author}{\bibfnamefont{F.-J.} \bibnamefont{Decker}}, \bibinfo{author}{\bibfnamefont{M.~J.} \bibnamefont{Hogan}}, \bibinfo{author}{\bibfnamefont{C.}~\bibnamefont{Huang}}, \bibinfo{author}{\bibfnamefont{R.}~\bibnamefont{Iverson}}, \bibinfo{author}{\bibfnamefont{C.}~\bibnamefont{Joshi}}, \bibinfo{author}{\bibfnamefont{T.~C.} \bibnamefont{Katsouleas}}, \bibinfo{author}{\bibfnamefont{W.}~\bibnamefont{Lu}}, \bibnamefont{et~al.}, \bibinfo{journal}{Physical Review Letters} \textbf{\bibinfo{volume}{90}} (\bibinfo{year}{2003}), \urlprefix\url{https://doi.org/10.1103/physrevlett.90.214801}.

\bibitem[{\citenamefont{Muggli et~al.}(2008)\citenamefont{Muggli, Blue, Clayton, Decker, Hogan, Huang, Joshi, Katsouleas, Lu, Mori et~al.}}]{Muggli2008}
\bibinfo{author}{\bibfnamefont{P.}~\bibnamefont{Muggli}}, \bibinfo{author}{\bibfnamefont{B.~E.} \bibnamefont{Blue}}, \bibinfo{author}{\bibfnamefont{C.~E.} \bibnamefont{Clayton}}, \bibinfo{author}{\bibfnamefont{F.~J.} \bibnamefont{Decker}}, \bibinfo{author}{\bibfnamefont{M.~J.} \bibnamefont{Hogan}}, \bibinfo{author}{\bibfnamefont{C.}~\bibnamefont{Huang}}, \bibinfo{author}{\bibfnamefont{C.}~\bibnamefont{Joshi}}, \bibinfo{author}{\bibfnamefont{T.~C.} \bibnamefont{Katsouleas}}, \bibinfo{author}{\bibfnamefont{W.}~\bibnamefont{Lu}}, \bibinfo{author}{\bibfnamefont{W.~B.} \bibnamefont{Mori}}, \bibnamefont{et~al.}, \bibinfo{journal}{Physical Review Letters} \textbf{\bibinfo{volume}{101}} (\bibinfo{year}{2008}), \urlprefix\url{https://doi.org/10.1103/physrevlett.101.055001}.

\bibitem[{\citenamefont{Corde et~al.}(2015)\citenamefont{Corde, Adli, Allen, An, Clarke, Clayton, Delahaye, Frederico, Gessner, Green et~al.}}]{Corde2015}
\bibinfo{author}{\bibfnamefont{S.}~\bibnamefont{Corde}}, \bibinfo{author}{\bibfnamefont{E.}~\bibnamefont{Adli}}, \bibinfo{author}{\bibfnamefont{J.~M.} \bibnamefont{Allen}}, \bibinfo{author}{\bibfnamefont{W.}~\bibnamefont{An}}, \bibinfo{author}{\bibfnamefont{C.~I.} \bibnamefont{Clarke}}, \bibinfo{author}{\bibfnamefont{C.~E.} \bibnamefont{Clayton}}, \bibinfo{author}{\bibfnamefont{J.~P.} \bibnamefont{Delahaye}}, \bibinfo{author}{\bibfnamefont{J.}~\bibnamefont{Frederico}}, \bibinfo{author}{\bibfnamefont{S.}~\bibnamefont{Gessner}}, \bibinfo{author}{\bibfnamefont{S.~Z.} \bibnamefont{Green}}, \bibnamefont{et~al.}, \bibinfo{journal}{Nature} \textbf{\bibinfo{volume}{524}}, \bibinfo{pages}{442} (\bibinfo{year}{2015}), \urlprefix\url{https://doi.org/10.1038/nature14890}.

\bibitem[{\citenamefont{Doche et~al.}(2017)\citenamefont{Doche, Beekman, Corde, Allen, Clarke, Frederico, Gessner, Green, Hogan, O'Shea et~al.}}]{Doche2017}
\bibinfo{author}{\bibfnamefont{A.}~\bibnamefont{Doche}}, \bibinfo{author}{\bibfnamefont{C.}~\bibnamefont{Beekman}}, \bibinfo{author}{\bibfnamefont{S.}~\bibnamefont{Corde}}, \bibinfo{author}{\bibfnamefont{J.~M.} \bibnamefont{Allen}}, \bibinfo{author}{\bibfnamefont{C.~I.} \bibnamefont{Clarke}}, \bibinfo{author}{\bibfnamefont{J.}~\bibnamefont{Frederico}}, \bibinfo{author}{\bibfnamefont{S.~J.} \bibnamefont{Gessner}}, \bibinfo{author}{\bibfnamefont{S.~Z.} \bibnamefont{Green}}, \bibinfo{author}{\bibfnamefont{M.~J.} \bibnamefont{Hogan}}, \bibinfo{author}{\bibfnamefont{B.}~\bibnamefont{O'Shea}}, \bibnamefont{et~al.}, \bibinfo{journal}{Scientific Reports} \textbf{\bibinfo{volume}{7}} (\bibinfo{year}{2017}), \urlprefix\url{https://doi.org/10.1038/s41598-017-14524-4}.

\bibitem[{\citenamefont{Hue et~al.}(2021)\citenamefont{Hue, Cao, Andriyash, Knetsch, Hogan, Adli, Gessner, and Corde}}]{Hue2021}
\bibinfo{author}{\bibfnamefont{C.~S.} \bibnamefont{Hue}}, \bibinfo{author}{\bibfnamefont{G.~J.} \bibnamefont{Cao}}, \bibinfo{author}{\bibfnamefont{I.~A.} \bibnamefont{Andriyash}}, \bibinfo{author}{\bibfnamefont{A.}~\bibnamefont{Knetsch}}, \bibinfo{author}{\bibfnamefont{M.~J.} \bibnamefont{Hogan}}, \bibinfo{author}{\bibfnamefont{E.}~\bibnamefont{Adli}}, \bibinfo{author}{\bibfnamefont{S.}~\bibnamefont{Gessner}}, \bibnamefont{and} \bibinfo{author}{\bibfnamefont{S.}~\bibnamefont{Corde}}, \bibinfo{journal}{Physical Review Research} \textbf{\bibinfo{volume}{3}} (\bibinfo{year}{2021}), \urlprefix\url{https://doi.org/10.1103/physrevresearch.3.043063}.

\bibitem[{\citenamefont{Chiou et~al.}(1995)\citenamefont{Chiou, Katsouleas, Decker, Mori, Wurtele, Shvets, and Su}}]{Chiou1995}
\bibinfo{author}{\bibfnamefont{T.~C.} \bibnamefont{Chiou}}, \bibinfo{author}{\bibfnamefont{T.}~\bibnamefont{Katsouleas}}, \bibinfo{author}{\bibfnamefont{C.}~\bibnamefont{Decker}}, \bibinfo{author}{\bibfnamefont{W.~B.} \bibnamefont{Mori}}, \bibinfo{author}{\bibfnamefont{J.~S.} \bibnamefont{Wurtele}}, \bibinfo{author}{\bibfnamefont{G.}~\bibnamefont{Shvets}}, \bibnamefont{and} \bibinfo{author}{\bibfnamefont{J.~J.} \bibnamefont{Su}}, \bibinfo{journal}{Physics of Plasmas} \textbf{\bibinfo{volume}{2}}, \bibinfo{pages}{310} (\bibinfo{year}{1995}), \urlprefix\url{https://doi.org/10.1063/1.871107}.

\bibitem[{\citenamefont{Chiou and Katsouleas}(1998)}]{Chiou1998}
\bibinfo{author}{\bibfnamefont{T.~C.} \bibnamefont{Chiou}} \bibnamefont{and} \bibinfo{author}{\bibfnamefont{T.}~\bibnamefont{Katsouleas}}, \bibinfo{journal}{Physical Review Letters} \textbf{\bibinfo{volume}{81}}, \bibinfo{pages}{3411} (\bibinfo{year}{1998}), \urlprefix\url{https://doi.org/10.1103/physrevlett.81.3411}.

\bibitem[{\citenamefont{Schroeder et~al.}(1999)\citenamefont{Schroeder, Whittum, and Wurtele}}]{Schroeder1999}
\bibinfo{author}{\bibfnamefont{C.~B.} \bibnamefont{Schroeder}}, \bibinfo{author}{\bibfnamefont{D.~H.} \bibnamefont{Whittum}}, \bibnamefont{and} \bibinfo{author}{\bibfnamefont{J.~S.} \bibnamefont{Wurtele}}, \bibinfo{journal}{Physical Review Letters} \textbf{\bibinfo{volume}{82}}, \bibinfo{pages}{1177} (\bibinfo{year}{1999}), \urlprefix\url{https://doi.org/10.1103/physrevlett.82.1177}.

\bibitem[{\citenamefont{Lindstr{\o}m et~al.}(2018)\citenamefont{Lindstr{\o}m, Adli, Allen, An, Beekman, Clarke, Clayton, Corde, Doche, Frederico et~al.}}]{Lindstrm2018}
\bibinfo{author}{\bibfnamefont{C.}~\bibnamefont{Lindstr{\o}m}}, \bibinfo{author}{\bibfnamefont{E.}~\bibnamefont{Adli}}, \bibinfo{author}{\bibfnamefont{J.}~\bibnamefont{Allen}}, \bibinfo{author}{\bibfnamefont{W.}~\bibnamefont{An}}, \bibinfo{author}{\bibfnamefont{C.}~\bibnamefont{Beekman}}, \bibinfo{author}{\bibfnamefont{C.}~\bibnamefont{Clarke}}, \bibinfo{author}{\bibfnamefont{C.}~\bibnamefont{Clayton}}, \bibinfo{author}{\bibfnamefont{S.}~\bibnamefont{Corde}}, \bibinfo{author}{\bibfnamefont{A.}~\bibnamefont{Doche}}, \bibinfo{author}{\bibfnamefont{J.}~\bibnamefont{Frederico}}, \bibnamefont{et~al.}, \bibinfo{journal}{Physical Review Letters} \textbf{\bibinfo{volume}{120}} (\bibinfo{year}{2018}), \urlprefix\url{https://doi.org/10.1103/physrevlett.120.124802}.

\bibitem[{\citenamefont{Gessner et~al.}(2016)\citenamefont{Gessner, Adli, Allen, An, Clarke, Clayton, Corde, Delahaye, Frederico, Green et~al.}}]{Gessner2016}
\bibinfo{author}{\bibfnamefont{S.}~\bibnamefont{Gessner}}, \bibinfo{author}{\bibfnamefont{E.}~\bibnamefont{Adli}}, \bibinfo{author}{\bibfnamefont{J.~M.} \bibnamefont{Allen}}, \bibinfo{author}{\bibfnamefont{W.}~\bibnamefont{An}}, \bibinfo{author}{\bibfnamefont{C.~I.} \bibnamefont{Clarke}}, \bibinfo{author}{\bibfnamefont{C.~E.} \bibnamefont{Clayton}}, \bibinfo{author}{\bibfnamefont{S.}~\bibnamefont{Corde}}, \bibinfo{author}{\bibfnamefont{J.~P.} \bibnamefont{Delahaye}}, \bibinfo{author}{\bibfnamefont{J.}~\bibnamefont{Frederico}}, \bibinfo{author}{\bibfnamefont{S.~Z.} \bibnamefont{Green}}, \bibnamefont{et~al.}, \bibinfo{journal}{Nature Communications} \textbf{\bibinfo{volume}{7}} (\bibinfo{year}{2016}), \urlprefix\url{https://doi.org/10.1038/ncomms11785}.

\bibitem[{\citenamefont{Muggli et~al.}(1999)\citenamefont{Muggli, Marsh, Wang, Clayton, Lee, Katsouleas, and Joshi}}]{Muggli1999}
\bibinfo{author}{\bibfnamefont{P.}~\bibnamefont{Muggli}}, \bibinfo{author}{\bibfnamefont{K.}~\bibnamefont{Marsh}}, \bibinfo{author}{\bibfnamefont{S.}~\bibnamefont{Wang}}, \bibinfo{author}{\bibfnamefont{C.}~\bibnamefont{Clayton}}, \bibinfo{author}{\bibfnamefont{S.}~\bibnamefont{Lee}}, \bibinfo{author}{\bibfnamefont{T.}~\bibnamefont{Katsouleas}}, \bibnamefont{and} \bibinfo{author}{\bibfnamefont{C.}~\bibnamefont{Joshi}}, \bibinfo{journal}{{IEEE} Transactions on Plasma Science} \textbf{\bibinfo{volume}{27}}, \bibinfo{pages}{791} (\bibinfo{year}{1999}), \urlprefix\url{https://doi.org/10.1109/27.774685}.

\bibitem[{\citenamefont{Kimura et~al.}(2011)\citenamefont{Kimura, Milchberg, Muggli, Li, and Mori}}]{Kimura2011}
\bibinfo{author}{\bibfnamefont{W.~D.} \bibnamefont{Kimura}}, \bibinfo{author}{\bibfnamefont{H.~M.} \bibnamefont{Milchberg}}, \bibinfo{author}{\bibfnamefont{P.}~\bibnamefont{Muggli}}, \bibinfo{author}{\bibfnamefont{X.}~\bibnamefont{Li}}, \bibnamefont{and} \bibinfo{author}{\bibfnamefont{W.~B.} \bibnamefont{Mori}}, \bibinfo{journal}{Physical Review Special Topics - Accelerators and Beams} \textbf{\bibinfo{volume}{14}} (\bibinfo{year}{2011}), \urlprefix\url{https://doi.org/10.1103/physrevstab.14.041301}.

\bibitem[{\citenamefont{Schr\"{o}der et~al.}(2020)\citenamefont{Schr\"{o}der, Lindstrøm, Bohlen, Boyle, D’Arcy, Diederichs, Garland, Gonzalez, Knetsch, Libov et~al.}}]{Schrder2020}
\bibinfo{author}{\bibfnamefont{S.}~\bibnamefont{Schr\"{o}der}}, \bibinfo{author}{\bibfnamefont{C.~A.} \bibnamefont{Lindstrøm}}, \bibinfo{author}{\bibfnamefont{S.}~\bibnamefont{Bohlen}}, \bibinfo{author}{\bibfnamefont{G.}~\bibnamefont{Boyle}}, \bibinfo{author}{\bibfnamefont{R.}~\bibnamefont{D’Arcy}}, \bibinfo{author}{\bibfnamefont{S.}~\bibnamefont{Diederichs}}, \bibinfo{author}{\bibfnamefont{M.~J.} \bibnamefont{Garland}}, \bibinfo{author}{\bibfnamefont{P.}~\bibnamefont{Gonzalez}}, \bibinfo{author}{\bibfnamefont{A.}~\bibnamefont{Knetsch}}, \bibinfo{author}{\bibfnamefont{V.}~\bibnamefont{Libov}}, \bibnamefont{et~al.}, \bibinfo{journal}{Nature Communications} \textbf{\bibinfo{volume}{11}} (\bibinfo{year}{2020}), ISSN \bibinfo{issn}{2041-1723}, \urlprefix\url{http://dx.doi.org/10.1038/s41467-020-19811-9}.

\bibitem[{\citenamefont{Hogan et~al.}(2010)\citenamefont{Hogan, Raubenheimer, Seryi, Muggli, Katsouleas, Huang, Lu, An, Marsh, Mori et~al.}}]{Hogan2010}
\bibinfo{author}{\bibfnamefont{M.~J.} \bibnamefont{Hogan}}, \bibinfo{author}{\bibfnamefont{T.~O.} \bibnamefont{Raubenheimer}}, \bibinfo{author}{\bibfnamefont{A.}~\bibnamefont{Seryi}}, \bibinfo{author}{\bibfnamefont{P.}~\bibnamefont{Muggli}}, \bibinfo{author}{\bibfnamefont{T.}~\bibnamefont{Katsouleas}}, \bibinfo{author}{\bibfnamefont{C.}~\bibnamefont{Huang}}, \bibinfo{author}{\bibfnamefont{W.}~\bibnamefont{Lu}}, \bibinfo{author}{\bibfnamefont{W.}~\bibnamefont{An}}, \bibinfo{author}{\bibfnamefont{K.~A.} \bibnamefont{Marsh}}, \bibinfo{author}{\bibfnamefont{W.~B.} \bibnamefont{Mori}}, \bibnamefont{et~al.}, \bibinfo{journal}{New Journal of Physics} \textbf{\bibinfo{volume}{12}}, \bibinfo{pages}{055030} (\bibinfo{year}{2010}), \urlprefix\url{https://doi.org/10.1088/1367-2630/12/5/055030}.

\bibitem[{\citenamefont{Berden et~al.}(2007)\citenamefont{Berden, Gillespie, Jamison, Knabbe, MacLeod, van~der Meer, Phillips, Schlarb, Schmidt, Schm{\"{u}}ser et~al.}}]{Berden2007}
\bibinfo{author}{\bibfnamefont{G.}~\bibnamefont{Berden}}, \bibinfo{author}{\bibfnamefont{W.~A.} \bibnamefont{Gillespie}}, \bibinfo{author}{\bibfnamefont{S.~P.} \bibnamefont{Jamison}}, \bibinfo{author}{\bibfnamefont{E.-A.} \bibnamefont{Knabbe}}, \bibinfo{author}{\bibfnamefont{A.~M.} \bibnamefont{MacLeod}}, \bibinfo{author}{\bibfnamefont{A.~F.~G.} \bibnamefont{van~der Meer}}, \bibinfo{author}{\bibfnamefont{P.~J.} \bibnamefont{Phillips}}, \bibinfo{author}{\bibfnamefont{H.}~\bibnamefont{Schlarb}}, \bibinfo{author}{\bibfnamefont{B.}~\bibnamefont{Schmidt}}, \bibinfo{author}{\bibfnamefont{P.}~\bibnamefont{Schm{\"{u}}ser}}, \bibnamefont{et~al.}, \bibinfo{journal}{Physical Review Letters} \textbf{\bibinfo{volume}{99}}, \bibinfo{pages}{164801} (\bibinfo{year}{2007}), ISSN \bibinfo{issn}{0031-9007}, \urlprefix\url{http://link.aps.org/doi/10.1103/PhysRevLett.99.164801}.

\bibitem[{\citenamefont{Andreev et~al.}(1996)\citenamefont{Andreev, Bychkov, Kotlyar, Margolin, Pyatnitskii, and Serafimovich}}]{Andreev1996}
\bibinfo{author}{\bibfnamefont{N.~E.} \bibnamefont{Andreev}}, \bibinfo{author}{\bibfnamefont{S.~S.} \bibnamefont{Bychkov}}, \bibinfo{author}{\bibfnamefont{V.~V.} \bibnamefont{Kotlyar}}, \bibinfo{author}{\bibfnamefont{L.~Y.} \bibnamefont{Margolin}}, \bibinfo{author}{\bibfnamefont{L.~N.} \bibnamefont{Pyatnitskii}}, \bibnamefont{and} \bibinfo{author}{\bibfnamefont{P.~G.} \bibnamefont{Serafimovich}}, \bibinfo{journal}{Quantum Electronics} \textbf{\bibinfo{volume}{26}}, \bibinfo{pages}{126} (\bibinfo{year}{1996}), \urlprefix\url{https://doi.org/10.1070/qe1996v026n02abeh000607}.

\bibitem[{\citenamefont{Fan et~al.}(2000)\citenamefont{Fan, Parra, Alexeev, Kim, Milchberg, Margolin, and Pyatnitskii}}]{Fan2000}
\bibinfo{author}{\bibfnamefont{J.}~\bibnamefont{Fan}}, \bibinfo{author}{\bibfnamefont{E.}~\bibnamefont{Parra}}, \bibinfo{author}{\bibfnamefont{I.}~\bibnamefont{Alexeev}}, \bibinfo{author}{\bibfnamefont{K.~Y.} \bibnamefont{Kim}}, \bibinfo{author}{\bibfnamefont{H.~M.} \bibnamefont{Milchberg}}, \bibinfo{author}{\bibfnamefont{L.~Y.} \bibnamefont{Margolin}}, \bibnamefont{and} \bibinfo{author}{\bibfnamefont{L.~N.} \bibnamefont{Pyatnitskii}}, \bibinfo{journal}{Physical Review E} \textbf{\bibinfo{volume}{62}}, \bibinfo{pages}{R7603} (\bibinfo{year}{2000}), \urlprefix\url{https://doi.org/10.1103/physreve.62.r7603}.

\bibitem[{\citenamefont{Clayton et~al.}(2016)\citenamefont{Clayton, Adli, Allen, An, Clarke, Corde, Frederico, Gessner, Green, Hogan et~al.}}]{Clayton2016}
\bibinfo{author}{\bibfnamefont{C.~E.} \bibnamefont{Clayton}}, \bibinfo{author}{\bibfnamefont{E.}~\bibnamefont{Adli}}, \bibinfo{author}{\bibfnamefont{J.}~\bibnamefont{Allen}}, \bibinfo{author}{\bibfnamefont{W.}~\bibnamefont{An}}, \bibinfo{author}{\bibfnamefont{C.~I.} \bibnamefont{Clarke}}, \bibinfo{author}{\bibfnamefont{S.}~\bibnamefont{Corde}}, \bibinfo{author}{\bibfnamefont{J.}~\bibnamefont{Frederico}}, \bibinfo{author}{\bibfnamefont{S.}~\bibnamefont{Gessner}}, \bibinfo{author}{\bibfnamefont{S.~Z.} \bibnamefont{Green}}, \bibinfo{author}{\bibfnamefont{M.~J.} \bibnamefont{Hogan}}, \bibnamefont{et~al.}, \bibinfo{journal}{Nature Communications} \textbf{\bibinfo{volume}{7}} (\bibinfo{year}{2016}), \urlprefix\url{https://doi.org/10.1038/ncomms12483}.

\bibitem[{\citenamefont{Gessner}(2016)}]{Gessner2016Thesis}
\bibinfo{author}{\bibfnamefont{S.~J.} \bibnamefont{Gessner}}, Ph.D. thesis, \bibinfo{school}{Stanford University} (\bibinfo{year}{2016}), \urlprefix\url{https://doi.org/10.2172/1340170}.

\bibitem[{\citenamefont{{Perelomov} et~al.}(1966)\citenamefont{{Perelomov}, {Popov}, and {Terent'ev}}}]{PPT1966}
\bibinfo{author}{\bibfnamefont{A.~M.} \bibnamefont{{Perelomov}}}, \bibinfo{author}{\bibfnamefont{V.~S.} \bibnamefont{{Popov}}}, \bibnamefont{and} \bibinfo{author}{\bibfnamefont{M.~V.} \bibnamefont{{Terent'ev}}}, \bibinfo{journal}{Soviet Journal of Experimental and Theoretical Physics} \textbf{\bibinfo{volume}{23}}, \bibinfo{pages}{924} (\bibinfo{year}{1966}).

\bibitem[{\citenamefont{Schroeder et~al.}(2013)\citenamefont{Schroeder, Esarey, Benedetti, and Leemans}}]{Schroeder2013}
\bibinfo{author}{\bibfnamefont{C.~B.} \bibnamefont{Schroeder}}, \bibinfo{author}{\bibfnamefont{E.}~\bibnamefont{Esarey}}, \bibinfo{author}{\bibfnamefont{C.}~\bibnamefont{Benedetti}}, \bibnamefont{and} \bibinfo{author}{\bibfnamefont{W.~P.} \bibnamefont{Leemans}}, \bibinfo{journal}{Physics of Plasmas} \textbf{\bibinfo{volume}{20}}, \bibinfo{pages}{080701} (\bibinfo{year}{2013}), \urlprefix\url{https://doi.org/10.1063/1.4817799}.

\bibitem[{\citenamefont{Shvets et~al.}(1996)\citenamefont{Shvets, Wurtele, Chiou, and Katsouleas}}]{Shvets1996}
\bibinfo{author}{\bibfnamefont{G.}~\bibnamefont{Shvets}}, \bibinfo{author}{\bibfnamefont{J.}~\bibnamefont{Wurtele}}, \bibinfo{author}{\bibfnamefont{T.}~\bibnamefont{Chiou}}, \bibnamefont{and} \bibinfo{author}{\bibfnamefont{T.}~\bibnamefont{Katsouleas}}, \bibinfo{journal}{{IEEE} Transactions on Plasma Science} \textbf{\bibinfo{volume}{24}}, \bibinfo{pages}{351} (\bibinfo{year}{1996}), \urlprefix\url{https://doi.org/10.1109/27.509999}.

\bibitem[{\citenamefont{Feder et~al.}(2020)\citenamefont{Feder, Miao, Shrock, Goffin, and Milchberg}}]{Feder2020}
\bibinfo{author}{\bibfnamefont{L.}~\bibnamefont{Feder}}, \bibinfo{author}{\bibfnamefont{B.}~\bibnamefont{Miao}}, \bibinfo{author}{\bibfnamefont{J.~E.} \bibnamefont{Shrock}}, \bibinfo{author}{\bibfnamefont{A.}~\bibnamefont{Goffin}}, \bibnamefont{and} \bibinfo{author}{\bibfnamefont{H.~M.} \bibnamefont{Milchberg}}, \bibinfo{journal}{Physical Review Research} \textbf{\bibinfo{volume}{2}} (\bibinfo{year}{2020}), \urlprefix\url{https://doi.org/10.1103/physrevresearch.2.043173}.

\bibitem[{\citenamefont{Yakimenko et~al.}(2019)\citenamefont{Yakimenko, Alsberg, Bong, Bouchard, Clarke, Emma, Green, Hast, Hogan, Seabury et~al.}}]{Yakimenko2019}
\bibinfo{author}{\bibfnamefont{V.}~\bibnamefont{Yakimenko}}, \bibinfo{author}{\bibfnamefont{L.}~\bibnamefont{Alsberg}}, \bibinfo{author}{\bibfnamefont{E.}~\bibnamefont{Bong}}, \bibinfo{author}{\bibfnamefont{G.}~\bibnamefont{Bouchard}}, \bibinfo{author}{\bibfnamefont{C.}~\bibnamefont{Clarke}}, \bibinfo{author}{\bibfnamefont{C.}~\bibnamefont{Emma}}, \bibinfo{author}{\bibfnamefont{S.}~\bibnamefont{Green}}, \bibinfo{author}{\bibfnamefont{C.}~\bibnamefont{Hast}}, \bibinfo{author}{\bibfnamefont{M.}~\bibnamefont{Hogan}}, \bibinfo{author}{\bibfnamefont{J.}~\bibnamefont{Seabury}}, \bibnamefont{et~al.}, \bibinfo{journal}{Physical Review Accelerators and Beams} \textbf{\bibinfo{volume}{22}} (\bibinfo{year}{2019}), \urlprefix\url{https://doi.org/10.1103/physrevaccelbeams.22.101301}.

\bibitem[{\citenamefont{Zhou et~al.}(2021)\citenamefont{Zhou, Hua, An, Mori, Joshi, Gao, and Lu}}]{Zhou2021}
\bibinfo{author}{\bibfnamefont{S.}~\bibnamefont{Zhou}}, \bibinfo{author}{\bibfnamefont{J.}~\bibnamefont{Hua}}, \bibinfo{author}{\bibfnamefont{W.}~\bibnamefont{An}}, \bibinfo{author}{\bibfnamefont{W.~B.} \bibnamefont{Mori}}, \bibinfo{author}{\bibfnamefont{C.}~\bibnamefont{Joshi}}, \bibinfo{author}{\bibfnamefont{J.}~\bibnamefont{Gao}}, \bibnamefont{and} \bibinfo{author}{\bibfnamefont{W.}~\bibnamefont{Lu}}, \bibinfo{journal}{Physical Review Letters} \textbf{\bibinfo{volume}{127}} (\bibinfo{year}{2021}), \urlprefix\url{https://doi.org/10.1103/physrevlett.127.174801}.

\bibitem[{\citenamefont{Clendenin et~al.}(1988)}]{Clendenin:1988np}
\bibinfo{author}{\bibfnamefont{J.~E.} \bibnamefont{Clendenin}} \bibnamefont{et~al.}, in \emph{\bibinfo{booktitle}{{14th International Linear Accelerator Conference}}} (\bibinfo{year}{1988}).

\bibitem[{\citenamefont{An et~al.}(2013)\citenamefont{An, Decyk, Mori, and Antonsen}}]{An2013}
\bibinfo{author}{\bibfnamefont{W.}~\bibnamefont{An}}, \bibinfo{author}{\bibfnamefont{V.~K.} \bibnamefont{Decyk}}, \bibinfo{author}{\bibfnamefont{W.~B.} \bibnamefont{Mori}}, \bibnamefont{and} \bibinfo{author}{\bibfnamefont{T.~M.} \bibnamefont{Antonsen}}, \bibinfo{journal}{Journal of Computational Physics} \textbf{\bibinfo{volume}{250}}, \bibinfo{pages}{165} (\bibinfo{year}{2013}), \urlprefix\url{https://doi.org/10.1016/j.jcp.2013.05.020}.

\end{thebibliography}

\end{document}